\documentclass[11pt,a4paper,final]{article}

\usepackage[a4paper,left=3cm,right=3cm,top=3cm,bottom=3cm]{geometry}

\usepackage[utf8]{inputenc}
\usepackage[USenglish]{babel}

\usepackage{amsmath}
\usepackage{amsfonts}
\usepackage{amssymb}
\usepackage{mathtools}
\usepackage{suffix}
\usepackage{graphicx}
\usepackage{xcolor}
\usepackage{enumerate}
\usepackage{setspace}

\usepackage{booktabs}
\usepackage{longtable}
\usepackage{array}
\usepackage{calc}

\usepackage{hyperref}
\hypersetup{
	pdftitle={Constraints on the String T-Duality Propagator from the Hydrogen Atom},
	pdfauthor={Michael Florian Wondrak},
	pdffitwindow=true,
	colorlinks=true,
	linkcolor={red!50!black},
	citecolor={blue!50!black},
	urlcolor={blue!80!black}
}

\usepackage[comma,numbers,sort&compress]{natbib}
\usepackage{siunitx}
\DeclareSIUnit\fm{\femto\meter}
\DeclareSIUnit\am{\atto\meter}

\newcommand{\inm}[1]{\ensuremath{\text{#1}}}
\newcommand{\eu}{\ensuremath{ {\text{e}} }}
\newcommand{\eulerMascheroni}{\ensuremath{ \gamma }}
\newcommand{\HStruve}[1]{\ensuremath{ \mathbf{H}_{#1} }}
\newcommand{\LAssLaguerre}[2]{\ensuremath{ L_{#1}^{#2} }}
\newcommand{\KBessel}[1]{\ensuremath{ K_{#1} }}
\newcommand{\YBessel}[1]{\ensuremath{ Y_{#1} }}
\DeclareMathOperator{\GamFct}{\Gamma}
\newcommand{\dext}{\ensuremath{ {\text{d}} }}
\newcommand{\der}{\ensuremath{ d }}
\newcommand{\del}{\ensuremath{ \partial }}
\DeclarePairedDelimiterX\MeijerM[3]{\lparen}{\rparen}%
{#3\, \delimsize\vert \begin{smallmatrix}#1 \\ #2\end{smallmatrix}}
\newcommand\MeijerG[8][]{%
  G^{\,#2,#3}_{#4,#5}\MeijerM[#1]{#6}{#7}{#8}}
\WithSuffix\newcommand\MeijerG*[7]{%
  G^{\,#1,#2}_{#3,#4}\MeijerM*{#5}{#6}{#7}}
\newcommand*{\Laplace}{\mathop{}\!\mathbin\Delta}

\newcommand{\lpl}{\ensuremath{ {l_\inm{P}} }}
\newcommand{\hPlanck}{\ensuremath{ h }}

\newcommand{\lambdaz}{\ensuremath{\lambda_\inm{0}}}
\newcommand{\DeltaE}[2]{\ensuremath{ \Delta E_{#1}^\inm{#2} }}
\newcommand{\Deltanu}[2]{\ensuremath{ \Delta \nu_{#1}^\inm{#2} }}
\newcommand{\indicesnu}[2]{\ensuremath{\nu_\inm{#1}^\inm{#2}}}
\newcommand{\intDummy}{\ensuremath{ y }}
\newcommand{\lz}{\ensuremath{ {l_\inm{0}} }}
\newcommand{\Rsd}{\ensuremath{ {R^\star} }}
\newcommand{\alphaP}{\ensuremath{ {\alpha'} }}
\newcommand{\indicesE}[2]{\ensuremath{E_\inm{#1}^\inm{#2}}}

\begin{document}

\begin{center}
\begin{spacing}{1.5}
{\Large\bf Constraints on the String T-Duality Propagator\\from the Hydrogen Atom}
\end{spacing}
\end{center}

\vspace{-0.1cm}
\begin{center}
\textbf{Michael~F.~Wondrak}$^{a,b,}$\footnote{\texttt{wondrak@fias.uni-frankfurt.de}}
\textbf{and Marcus~Bleicher}$^{a,b,}$\footnote{\texttt{bleicher@th.physik.uni-frankfurt.de}}

\vspace{.6truecm}
{\em $^a$Frankfurt Institute for Advandced Studies (FIAS),\\
Ruth-Moufang-Stra\ss{}e 1, 60438 Frankfurt am Main, Germany}\\

\vspace{.3truecm}
{\em $^b$Institut f\"ur Theoretische Physik,\\
Johann Wolfgang Goethe-Universit\"at Frankfurt am Main,\\
Max-von-Laue-Stra\ss{}e 1, 60438 Frankfurt am Main, Germany}\\

\vspace{.6truecm}
October 17, 2019
\end{center}

\vspace{0.1cm}

\begin{abstract}
\noindent{\small%
We investigate the implications of a string-theory modified propagator in the high-precision regime of quantum mechanics. In particular, we examine the situation in which string theory is compactified at the T-duality self-dual radius. The corresponding propagator is closely related to the one derived from the path integral duality.

Our focus is on the hydrogen ground state energy and the $1\text{S}_{1/2}-2\text{S}_{1/2}$ transition frequency as they are the most precisely explored properties of the hydrogen atom. In our analysis, the T-duality propagator affects the photon field leading to a modified Coulomb potential. Thus, our study is complementary to investigations where the electron evolution is modified as in studies of a minimal length in the context of the generalized uncertainty principle.

The first manifestation of the T-duality propagator arises at fourth order in the fine-structure constant, including a logarithmic term. The constraints on the underlying parameter, the zero-point length, reach down to $\SI{3.9e-19}{\m}$ and are in full agreement with previous studies on black holes.

\bigskip\par
{\em Keywords:} 
String T-duality, %
zero-point length, %
minimal length, %
extra dimensions, %
modified Coulomb potential,
hydrogen energy levels
}
\end{abstract}

\clearpage

\section{Introduction}
\label{sec:Intro}
Symmetries lie at the heart of almost any theory in physics and imply far-reaching consequences. For example, symmetries powerfully constrain the structure of terms that are allowed in a given action (cf.~the electroweak theory, e.g.~\cite{Weinberg2015bInbookElectroweakTheory}).  
They also allow to clearly identify the fundamental degrees of freedom by choosing in appropriate gauge (cf.~gravitational waves, e.g.~\cite{deRham2014,Maggiore2017InbookTTGauge}).
Furthermore, in the context of string theory, there are even symmetries which show the equivalence of whole theories. 
Among those is T-duality which by acting on the moduli space relates string theories compactified on different backgrounds. 

As a special case, T-duality relates toroidally compactified theories which emerge from each other under inversion of the compactification radius, $R$, if associated with the exchange of Kaluza-Klein mode numbers, $n$, with winding mode numbers, $w$.
In the case of one extra dimension, the relation reads $R \to \Rsd^2/R$ and $n \leftrightarrow w$. The self-dual radius $\Rsd$ is mapped onto itself while compactification radii smaller than $\Rsd$ are identified with larger ones. Thus a notion of a smallest sensible length scale arises.
The self-dual radius is located at the string scale, $\Rsd = \sqrt{\alphaP}$, where $\alphaP$ denotes the Regge slope.

Starting from toroidally compactified bosonic string theory with compactification radius $\Rsd$, the authors of \cite{FontaniniSP2006} derived an effective 4-dimensional propagator for the center-of-mass of closed strings (cf.~also \cite{SmailagicSP2003,SpallucciF2005}). Compared to standard quantum field propagators, the presence of the compactified extra dimensions implies a UV finite behavior. The propagator receives contributions from a tower of momentum and winding modes. Out of those, the massless mode is the most relevant for low-energy physics. It gives rise to the T-duality propagator.

A related concept is the path integral duality which implements a scale-inversion symmetry in quantum field theory at a scale given by the so-called zero-point length, $\lz$ \cite{Padmanabhan1997,Padmanabhan1998,SrinivasanSP1998}. The propagators obtained by the T-duality approach and by the path integral duality agree to first order. This enables us to identify $\lz = 2\pi \Rsd = 2\pi \sqrt{\alphaP}$.

Based on this concept, a black hole solution was derived recently \cite{NicoliniSW2019}. The zero-point length plays a crucial role since it leads to a resolution of the curvature singularity at the black hole center and offers a non-divergent black hole evaporation process without a final explosion. This provides a possible observable to test the T-duality approach. 

The aim of this paper is to study the concept of the T-duality induced zero-point length from a different, low-energy perspective. We choose a system in quantum physics which is investigated to high precision by theory and experiment -- the hydrogen atom \cite{MohrNT2015} -- and derive constraints on $\lz$. The hydrogen atom has been used by several authors to investigate high-energy concepts, e.g.~\cite{Brau1999,AkhouryY2003,HossenfelderBHRSS2003,AntonacciOakesFFN2013,WondrakNB2016}.
We use two characteristics of the hydrogen atom: firstly the ground state energy and secondly the transition frequency between the two lowest levels with vanishing orbital angular momentum, $1\inm{S}_{1/2}$ and $2\inm{S}_{1/2}$ -- the spectral line which is experimentally known to highest precision \cite{Kramida2010}. The potential shift in these observables due to the T-duality concept strongly depends on the size of $\lz$. We calculate those in Rayleigh-Schr\"odinger perturbation theory. From comparison with discrepancies between experimental and theoretical values and from their uncertainties, we obtain upper limits on $\lz$.

In Sec.~\ref{sec:Hydrogen_E_Levels}, we review the theoretical description of the hydrogen atom. Moreover, we obtain the energy and frequency contributions from considering the T-duality propagator. The constraints from both observables are derived in Sec.~\ref{sec:Constraints_l0} and are discussed and put into context in Sec.~\ref{sec:Discussion}. Section~\ref{sec:Summary} offers a summary. Useful mathematical identities are presented in Sec.~\ref{sec:Useful_Identities} in the appendix.
Throughout this article we use natural units, in particular $c \equiv \hbar \equiv 1$. For electromagnetic quantities we apply the Lorentz-Heaviside convention which additionally implies $\epsilon_0 \equiv \mu_0 \equiv 1$.

\section{Hydrogen Atom Energy Levels}
\label{sec:Hydrogen_E_Levels}
The hydrogen atom is a prime object in quantum mechanics. This section focuses on the energy spectrum in the conventional description and on the corrections arising from a T-self-dual spacetime. 
We start with the Schr\"odinger equation with fine-structure terms because we consider a low-energy quantum system. Then we introduce the T-duality induced modifications of the Coulomb potential. Finally, we derive shifts in energy levels and transition frequencies.

\subsection{Conventional Description}
The stationary Schr\"odinger equation, 
$H_0 \left|\psi\right\rangle = E \left|\psi\right\rangle$,
with the eigenvectors $\left|\psi\right\rangle$ and eigenvalues $E$, is the starting point of our recapitulation which follows the Refs.~\cite{GreinerIV2005InbookHydrogen} and \cite{SchwablQMI2007InbookHydrogen}. 
In position space, the Hamiltonian for spherically symmetric systems is given by
\begin{align}
H_0
= -\frac{1}{2\mu}\, \Laplace +V_0
= -\frac{1}{2\mu r^2}\, \del_r \left(r^2 \del_r\right) +\frac{\vec{L}^2}{2\mu r^2} +V_0.
\end{align}
Here, $\mu$ is the reduced mass of the electron-proton system, $\Laplace$ is the Laplace operator, and $\vec{L}$ the angular momentum operator. The potential term follows directly from the Coulomb interaction,
\begin{equation}
V_0
= -\frac{\alpha}{r},
\end{equation}
where $\alpha = e^2/4\pi$ denotes Sommerfeld's fine-structure constant. The well-known separation ansatz 
in spherical coordinates reads 
\begin{equation}
\psi_{nlm}^{(0)}\!\left(r,\vartheta,\varphi\right)
\equiv \left\langle\, \vec{r} \vphantom{{nlm}^{\left(0\right)}} \right. %
  \,\left|\, {nlm}^{\left(0\right)} \right\rangle
= \frac{u_{nl}\left(r\right)}{r}\, Y_{lm}\!\left(\vartheta,\varphi\right).
\end{equation}
with the principle, the orbital angular momentum, and the magnetic quantum numbers $n$, $l$, and $m$.
The spherical harmonics $Y_{lm}\!\left(\vartheta,\varphi\right)$ solve the angular part while the radial equation simplifies to
\begin{equation}
\left( 
  -\frac{1}{2\mu}\, \frac{\der^2}{\der r^2}
  +\frac{1}{2\mu}\, \frac{l \left(l+1\right)}{r^2}
  -\frac{\alpha}{r}
\right)
u_{nl}\!\left(r\right) 
= E_{n}\,u_{nl}\!\left(r\right)
\label{eq:SchEq}
\end{equation}
with the bound-state solutions
\begin{equation}
u_{nl}\!\left(r\right)
= -{\left[
	\frac{\left( n-l-1 \right)!\, {\left(2\kappa\right)}^3}%
	{2n\, {\left( \left(n+l\right)! \right)}^3}
 \right]}^{1/2}
 r\, {\left(2\kappa r\right)}^l\, \eu^{-\kappa r}\,
 \LAssLaguerre{n+l}{2l+1}\left(2\kappa r\right).
\end{equation}
Herein, we define $\kappa \equiv \mu \alpha/n$ and apply the associated Laguerre polynomials
\begin{equation}
\LAssLaguerre{r}{s}\left(x\right)
= \sum_{k=0}^{r-s} {\left(-1\right)}^{k+s}\, 
  \frac{{\left(r!\right)}^2}{k!\, {\left(k+s\right)}!\, {\left(r-k-s\right)}!}\, 
  x^k.
\end{equation}
The spectrum of the energy eigenstates is discrete and depends on $n$ only,
\begin{equation}
E_{n}
= -\frac{\mu\alpha^2}{2}\, \frac{1}{n^2}.
\end{equation}
The ground state, denoted by $1\inm{S}_{1/2}$, shows the energy $\indicesE{th}{S} \equiv E_1 = -\mu\alpha^2/2$. Below, we will also employ the first excited state of spherical symmetry, $2\inm{S}_{1/2}$. The associated wave functions read
\begin{align}
\psi_{100}^{(0)}
&= \frac{2}{\sqrt{4\pi}}\, {\left(\mu\alpha\right)}^{3/2}\, e^{-\alpha\mu r}\\
\psi_{200}^{(0)}
&= \frac{2}{\sqrt{4\pi}}\, {\left(\frac{\mu\alpha}{2}\right)}^{3/2}\, 
  \left[1 -\frac{1}{2}\, \alpha\mu r\right]\, e^{-\alpha\mu r/2},
\end{align}
and the transition frequency between both states follows to be 
$\indicesnu{th}{S} \equiv \left(E_2 -E_1\right)/2\pi$.

Next, we include relativistic corrections which lead to the so-called fine structure in the spectrum. These are encoded in the relativistically adjusted Hamiltonian $H = H_0 +H_\text{fs}$. The terms for relativistic momentum correction, spin-orbit coupling, and zitterbewegung (Darwin term) constitute the Hamiltonian contribution
\begin{equation}
H_\text{fs}
= -\frac{\Laplace^2}{8\mu^3} 
 +\frac{\alpha}{4\mu^2 r^3}\, \vec{\sigma} \cdot \vec{L}
 +\frac{1}{8\mu^2}\, \left(\Laplace V_0\right),
\end{equation}
which involves the Pauli matrices, $\vec{\sigma}$.

The correction in the energy spectrum can be calculated by means of the time-independent Rayleigh-Schr\"odinger perturbation theory. 
The first order corrections partially break the degeneracy and explicitly depend on the total angular momentum quantum number $j = l+s = l \pm 1/2$. They read 
\begin{align}
\DeltaE{n,j}{fs}
& = \left\langle {nlm}^{\left(0\right)} \,\right|\, 
    H_\text{fs} 
    \,\left|\, {nlm}^{\left(0\right)} \right\rangle \\
\label{eq:corrFS}
& = \frac{\mu \alpha^2}{2n^2}\, \frac{\alpha^2}{n^2}\, 
 \left(
	\frac{3}{4} -\frac{n}{j +1/2}
 \right).
\end{align}
In particular, the corrections to the $1\inm{S}_{1/2}$ and $2\inm{S}_{1/2}$ levels are
\begin{equation}
\DeltaE{1,1/2}{fs}
= -\frac{\mu \alpha^2}{2}\, \frac{\alpha^2}{4}, \qquad
\DeltaE{2,1/2}{fs}
= -\frac{\mu \alpha^2}{2}\, \frac{5\alpha^2}{128}.
\end{equation}
We define the improved value of the ground state energy as $\indicesE{th}{fs} \equiv E_1 + \DeltaE{1,1/2}{fs}$ and the corrected transition frequency as 
$\indicesnu{th}{fs} \equiv \left(\left(E_2 +\DeltaE{2,1/2}{fs}\right) -\left(E_1 +\DeltaE{1,1/2}{fs}\right)\right)/2\pi$.

The fine-structure corrections naturally arise in the Dirac treatment of the hydrogen atom and agree with eq.~\eqref{eq:corrFS} to order $\alpha^4$. The state-of-the-art description of the hydrogen atom goes beyond idealizations like that of a pointlike nucleus or vanishing nuclear polarizability, and uses methods of quantum field theory to include, e.g., multiple photon interactions in quantum electrodynamics or hadronic contributions to the proton self-energy from quantum chromodynamics. Those corrections appear at order $\alpha^5$ or higher. For an overview of the contributions, we refer the reader to \cite{MohrNT2015}.

\subsection{Contribution from T-Duality Propagator}
\label{subsec:T-Duality_Contribution}
The considerations up to now originate from quantum mechanics and quantum field theory. The Standard Model of particle physics, however, is generally assumed to be incomplete. In contrast, superstring theory is a possible candidate for a unified theory also valid at high energies which reduces to the Standard Model and to general relativity as limiting cases \cite{BlumenhagenCLS2005}. 
In the following we consider closed bosonic string theory on a manifold with toroidal compactification where the compactification radius equals the self-dual radius under T-duality. Regarding the 4-dimensional propagation, the string center of mass deviates from excitations of quantum fields. The Euclidean propagator of a massless scalar field inherited from bosonic string theory reads \cite{NicoliniSW2019}
\begin{equation}
G(k)
= -\frac{\lz}{\sqrt{k^2}}\, \KBessel{1} \!\left(\lz \sqrt{k^2}\right)
\to
\begin{cases}
-1/k^2 & \inm{if $k^2 \ll 1/\lz^2$}\\
-\lz^{1/2}\, {\left(k^2\right)}^{-3/4}\, e^{-\lz \sqrt{k^2}}& \inm{if $k^2 \gg 1/\lz^2$}
\end{cases} \,\, ,
\end{equation}
where $\KBessel{\nu}\!\left(x\right)$ are modified Bessel functions of the second kind. In the low-momentum limit, one obtains the standard scalar propagator, while there is an exponential suppression for momenta large compared to $1/\lz$.

\begin{figure*}[htbp]
\begin{center}
	\includegraphics[width=0.6\linewidth]{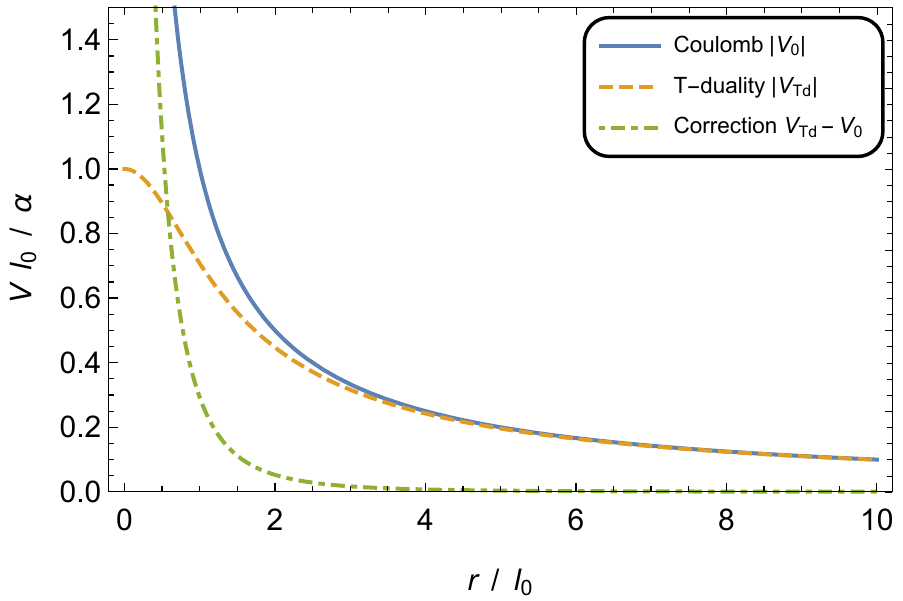}
\caption{Potential energy. The solid curve (blue) displays the absolute value of the conventional Coulomb energy, $\left|V_0\right|$, while the dashed line (orange) shows the absolute value of the T-duality corrected energy, $\left|V_\text{Td}\right|$. The difference of both equals the Hamiltonian contribution $H_\text{Td}$ (dot-dashed curve, green).}
\label{fig:plotpotentELegend}
\end{center}
\end{figure*}

If quantum fields behave this way, implications for the virtual-particle exchange can be expected. This leads to a modified interaction potential which explicitly includes the zero-point length as a UV cutoff \cite{NicoliniSW2019}. 
Applied to electrodynamics, the potential energy reads
\begin{equation}
V_\inm{Td}
= -\frac{\alpha}{\sqrt{r^2 +\lz^2}}.
\end{equation}
The difference to the conventional Coulomb interaction can be used to identify manifestations of T-duality from the hydrogen energy spectrum. From that we can derive constraints on $\lz$. Similar to the inclusion of fine-structure corrections, we apply the Rayleigh-Schr\"odinger perturbation theory to the amended Hamiltonian
$H = H_0 +H_\text{fs} +H_\text{Td}$. 
The additional term comprises the modification of the Coulomb energy,
\begin{equation}
H_\text{Td}
= V_\inm{Td} -V_0
= \frac{\alpha}{r} -\frac{\alpha}{\sqrt{r^2 +\lz^2}},
\end{equation}
which is presented in Fig.~\ref{fig:plotpotentELegend}.

In general, the level shifts depend on $n$ and $l$,
\begin{align}
\DeltaE{n,l}{Td}
&= \left\langle {nlm}^{\left(0\right)} \,\right|\, 
  H_\text{Td} 
  \,\left|\, {nlm}^{\left(0\right)} \right\rangle\\
\begin{split}
&= \frac{2^{2+2l} \left( n-l-1 \right)!\; \mu\alpha^2}{n^{4+2l} {\left( \left(n+l\right)! \right)}^3}\,\\
&\quad{} \times \int_0^\infty \dext \intDummy\; \intDummy^{2+2l}\, 
   \eu^{-2\intDummy/n}\,
   {\left[ \LAssLaguerre{n+l}{2l+1}\left(2\intDummy/n\right) \right]}^2
   \left(\frac{1}{\intDummy} -\frac{1}{\sqrt{\intDummy^2 +x^2}}\right).
\end{split}
\end{align}
Here we introduced $\intDummy \equiv \alpha\mu r$ and defined $x \equiv \lambdaz \alpha$ where $\lambdaz \equiv \mu \lz$.
For the ground state, we find the following expression at first order in perturbation theory:
\begin{align}
\label{eq:FullGrdState}
\DeltaE{1,0}{Td}
&= \frac{\mu \alpha^2}{2} \left(
	2 +\frac{16}{3}\, x^3
	+2\pi x \left[ \YBessel{1}\!\left(2x\right) +\HStruve{1}\!\left(2x\right) \right]
	-4\pi x^2 \left[ \YBessel{0}\!\left(2x\right) +\HStruve{2}\!\left(2x\right) \right]
 \right)\\
\label{eq:seriesGrdState}
& = \frac{\mu \alpha^2}{2} 
 \left(
	\left[ -2 -4\eulerMascheroni +4\ln\frac{1}{x} \right] x^2 
	+\mathcal{O}\left(x^3\right)
 \right)\\
\label{eq:seriesAlphaGrdState}
& = \mu \lambdaz^2 
 \left[ -1 -2\eulerMascheroni +2\ln\frac{1}{\alpha\lambdaz} \right]
 {\alpha}^4
 +\mathcal{O}\!\left(\alpha^5\right).
\end{align}
We used the Euler-Mascheroni constant, $\eulerMascheroni \approx \num{0.577}$, the Bessel functions of the second kind, $\YBessel{\nu}(x)$, and the Struve functions, $\HStruve{\nu}(x)$. 
The level shift of $2\inm{S}_{1/2}$ reads:
\begin{align}
\begin{split}
\DeltaE{2,0}{Td}
&= \frac{\mu \alpha^2}{2} \left(
	\frac{1}{2} +\frac{3}{4}\, x^3
	+\frac{\pi}{4} \left(x +x^3\right)
		 \left[ \YBessel{1}\!\left(x\right) -\HStruve{1}\!\left(x\right) \right]
 \right.\\		
&\quad \left.\vphantom{\frac{1}{2}}
	{}+\frac{\pi}{8} \left(-3x^2 +x^4\right)
		 \left[ \YBessel{0}\!\left(x\right) -\HStruve{0}\!\left(x\right) \right]	
 \right)
\end{split}\\
& = \frac{\mu \alpha^2}{2} 
 \left(
	\frac{1}{8}\left[ -5 -4\eulerMascheroni +4\ln\frac{2}{x} \right] x^2 
	+\mathcal{O}\left(x^3\right)
 \right)\\
& = \frac{\mu \lambdaz^2}{16}\, 
 \left[ -5 -4\eulerMascheroni +4\ln\frac{2}{\alpha\lambdaz} \right] {\alpha}^4
 +\mathcal{O}\!\left(\alpha^5\right).
\end{align}
We provide the identities crucial in deriving this result in Sec.~\ref{sec:Useful_Identities} in the appendix. Note that the corrections start at order $\alpha^4$. Note also that they are of the form $\DeltaE{}{} \propto \left(\inm{const.} +\ln\frac{1}{\alpha\mu\lz}\right) \lz^2$. For small values of $\lz$ the logarithmic term dominates ensuring the energy shifts to be positive. 
In contrast, the change in the transition frequency is negative for small values of $\lz$:
\begin{align}
\label{eq:FullTransFrequ}
\Deltanu{\inm{1S--2S}}{Td}
&= \frac{1}{2\pi} \left(\DeltaE{2,0}{Td} -\DeltaE{1,0}{Td}\right)\\
\label{eq:seriesTransFrequ}
& = \frac{1}{2\pi}\, \frac{\mu \alpha^2}{2} 
 \left(
	\frac{1}{8}\left[ 11 +4\ln 2 +28\eulerMascheroni -28\ln\frac{1}{x} \right] x^2 
	+\mathcal{O}\left(x^3\right)
 \right)\\
\label{eq:seriesAlphaTransFrequ}
& = \frac{\mu \lambdaz^2}{32\pi}\, 
 \left[ 11 +4\ln 2 +28\eulerMascheroni -28\ln\frac{1}{\alpha\lambdaz} \right]
 {\alpha}^4
 +\mathcal{O}\!\left(\alpha^5\right).
\end{align}

\section{Constraints on the Zero-Point Length}
\label{sec:Constraints_l0}
In the previous section, we derived the shifts in the energy levels $1\inm{S}_{1/2}$ and $2\inm{S}_{1/2}$ as well as the shift in the associated transition frequency as a function of the zero-point length. Now we can contrast the shifts with experimental data in order to obtain constraints on the value of $\lz$.

\subsection{Ground State Energy}
The reference values and uncertainties of the hydrogen ground state energy from theory and experiment are displayed in Tab.~\ref{tab:ThBackground_Energy_Comparison}.
Taking into account the respective standard deviations, we take the maximum differences between the fine-structure improved Schr\"odinger value, $\indicesE{th}{fs}$, and the experimental one, $\indicesE{exp}{}$ as well as between the current theoretical value, $\indicesE{th}{QED}$, and the experimental one. In this context, the experimental precision $\DeltaE{\inm{exp}}{}$ by itself defines the smallest upper bound on $\lz$.

\begin{table*}[!ht]
\centering
\begin{tabular}[l]{p{0.12\linewidth}p{0.40\linewidth}p{0.48\linewidth-6\tabcolsep}}
  \toprule
  {Energy} & {Description} & {Value}\\
  \midrule
  $\indicesE{th}{S}$
  &Schr\"odinger& $\SI{-13.59828715(9)}{\eV}$\\[3pt]
  $\indicesE{th}{fs}$
  &Schr\"odinger, incl.~fine-structure& $\SI{-13.59846818(9)}{\eV}$\\[3pt]
  $\indicesE{th}{QED}$
  &current theoretical value \cite{JentschuraKLMT2005}& $\SI{-13.59843449(9)}{\eV}$\\
  && $\SI{-3288086857.1276(31)}{\MHz} \cdot \hPlanck$ \\[3pt]
  $\indicesE{exp}{}$
  &current experimental value \cite{Kramida2010}&  $\SI{-13.59843448(9)}{\eV}$\\
  &&$\SI{-3288086856.8(7)}{\MHz} \cdot \hPlanck$\\[3pt]
  \bottomrule
\end{tabular}
\caption{Theoretical and experimental values of the hydrogen ground state energy. 
The calculation of $\indicesE{th}{S}$ and $\indicesE{th}{fs}$ is based on the 2014 CODATA recommended values \cite{MohrNT2015}.
When expressed in $\si{\eV}$, the actual precision of the current theoretical and experimental value is masked by the less precise known Planck constant \cite{MohrNT2015}. For this reason, $\hPlanck$ is factored out and the values are given also in terms of $\si{MHz}\cdot\hPlanck$. }
\label{tab:ThBackground_Energy_Comparison}
\end{table*}

\begin{figure*}[htbp]
\begin{center}
	\includegraphics[width=0.6\linewidth]{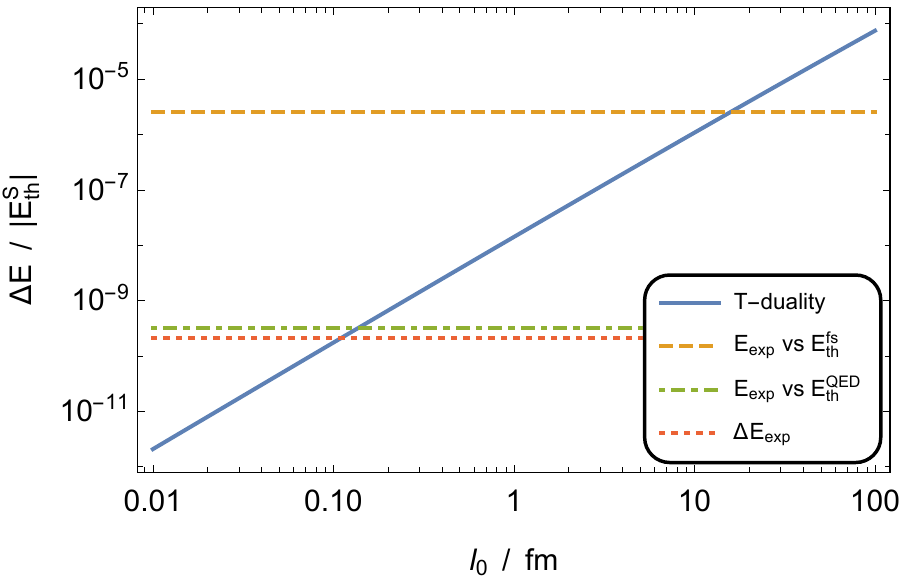}
\caption{Normalized uncertainty in the hydrogen ground state energy. The possible T-duality contribution $\DeltaE{1,0}{Td}$ (solid, blue) increases with the zero-point length, $\lz$. The dashed (orange) and dot-dashed (green) lines show the differences between the experimental value on the one hand and the fine-structure corrected or the current theoretical value on the other hand -- taking into account the respective standard deviations. The experimental error is presented by the dotted line (red).}
\label{fig:plotConstrGrdStateWLogLogLegend}
\end{center}
\end{figure*}

Figure~\ref{fig:plotConstrGrdStateWLogLogLegend} shows the relative T-duality contribution $\DeltaE{1,0}{Td}$ as a function of the zero-point length, $\lz$. The reference values are included as well. The zero-point length has to be smaller than the intersection point value to comply with the corresponding bound. We find the upper bounds for $\lz$ to be \SI{15.7}{\fm} (from comparison of the fine-structure improved Schr\"odinger description with the experiment), \SI{0.136}{\fm} (from comparison of the state-of-the-art theoretical value with the experiment), and \SI{0.112}{\fm} (from the experimental precision), respectively. For the sake of clarity and for contrasting with the other approach, the values are also presented in Tab.~\ref{tab:Constraints_Comparison}.

\subsection{Transition Frequency}
Among all transitions of states in the hydrogen atom, the transition frequency between the $1\inm{S}_{1/2}$ and the $2\inm{S}_{1/2}$ level is experimentally known with highest precision. The relative precision $\Deltanu{\inm{exp}}{}/\indicesnu{exp}{}$ of the experimental value surpasses the relative precision of the absolute ground state energy, $\DeltaE{\inm{exp}}{}/\indicesE{exp}{}$, by 5 orders of magnitude. Therefore we obtain more stringent upper bounds from the transition data than from the absolute energy data discussed above. The theoretical and experimental values of the transition frequencies associated with their uncertainties are presented in Tab.~\ref{tab:ThBackground_Frequ_Comparison}.

\begin{table*}[!ht]
\centering
\begin{tabular}[l]{p{0.12\linewidth}p{0.40\linewidth}p{0.48\linewidth-6\tabcolsep}}
  \toprule
  {Frequency} & {Description} & {Value}\\
  \midrule
  $\indicesnu{th}{S}$
  &Schr\"odinger& $\SI{2466038423(32)}{\MHz}$\\[3pt]
  $\indicesnu{th}{fs}$
  &Schr\"odinger, incl.~fine-structure& $\SI{2466068517(32)}{\MHz}$\\[3pt]
  $\indicesnu{th}{QED}$
  &current theoretical value \cite{JentschuraKLMT2005}& $\SI{2466061413.187103(46)}{\MHz}$\\[3pt]
  $\indicesnu{exp}{}$
  &current experimental value \cite{MatveevEtAl2013}
  &  $\SI{2466061413.187018(11)}{\MHz}$\\[3pt]
  \bottomrule
\end{tabular}
\caption{Theoretical and experimental values of the $1\text{S}_{1/2} - 2\text{S}_{1/2}$ hydrogen transition frequency. The calculation of $\indicesnu{th}{S}$ and $\indicesnu{th}{fs}$ is based on the 2014 CODATA recommended values \cite{MohrNT2015}.}
\label{tab:ThBackground_Frequ_Comparison}
\end{table*}

\begin{figure*}[htbp]
\begin{center}
	\includegraphics[width=0.6\linewidth]{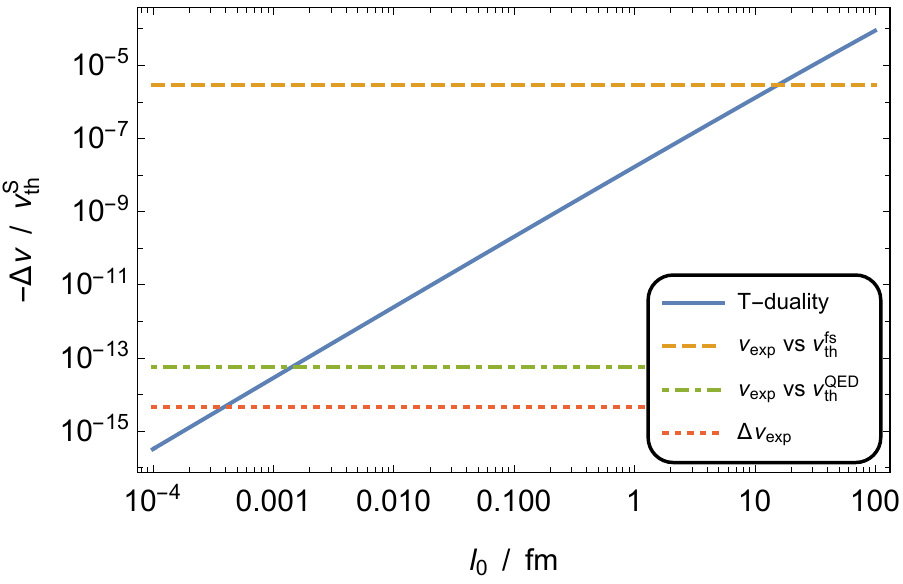}
\caption{Normalized uncertainty in the $1\text{S}_{1/2} - 2\text{S}_{1/2}$ hydrogen transition frequency. The possible T-duality contribution $\Deltanu{\inm{1S--2S}}{Td}$ (solid, blue) increases with the zero-point length, $\lz$. The dashed (orange) and dot-dashed (green) lines show the differences between the experimental value on the one hand and the fine-structure corrected or the current theoretical value on the other hand -- taking into account the respective standard deviations. The experimental error is presented by the dotted line (red).
}
\label{fig:plotConstrFrequWLogLogLegend}
\end{center}
\end{figure*}

Our approach is analogous as in the case of the ground state energy. In Fig.~\ref{fig:plotConstrFrequWLogLogLegend} one finds the $\lz$-dependent T-duality contribution to the transition frequency as well as the reference values.

The corresponding upper bounds on $\lz$ are
\SI{15.7}{\fm} (from comparison of the Schr\"odinger description including fine-structure corrections with the experiment), 
\SI{1.45e-3}{\fm} (from comparison of the current theoretical value with the experiment), and 
\SI{3.90e-4}{\fm} (from the experimental precision).
They are displayed in Tab.~\ref{tab:Constraints_Comparison}, opposed to their ground state energy counterparts.

\begin{table*}[ht]
\centering
\begin{tabular}[l]{p{0.20\linewidth}p{0.22\linewidth}||%
	p{0.20\linewidth}p{0.22\linewidth}}
  \toprule
  {Reference value} & {Upper bound on $\lz$} & {Reference value} & {Upper bound on $\lz$}\\
  \midrule
  $\indicesE{exp}{} -\indicesE{th}{fs}$ & \SI{1.6e-14}{\m}
  & $\indicesnu{exp}{} -\indicesnu{th}{fs}$ & \SI{1.6e-14}{\m}\\[3pt]
  $\indicesE{exp}{} -\indicesE{th}{QED}$ & \SI{1.4e-16}{\m}
  & $\indicesnu{exp}{} -\indicesnu{th}{QED}$ & \SI{1.5e-18}{\m}\\[3pt]
  $\DeltaE{\inm{exp}}{}$ & \SI{1.1e-16}{\m}
  & $\Deltanu{\inm{exp}}{}$ & \SI{3.9e-19}{\m}\\[3pt]
  \bottomrule
\end{tabular}
\caption{Bounds on $\lz$. This table summarizes the findings from Sec.~\ref{sec:Constraints_l0}.%
}
\label{tab:Constraints_Comparison}
\end{table*}

\section{Discussion}
\label{sec:Discussion}
Below, we discuss the validity and self-consistency of the results of Sec.~\ref{sec:Hydrogen_E_Levels} first. Then we comment on the bounds on the zero-point length.

One can classify the different contributions to the hydrogen energy levels in terms of powers of the fine-structure constant. Generally, the higher is the order of a term, the smaller is its absolute contribution. The Schr\"odinger value sets the scale at $\alpha^2$ and the Schr\"odinger fine-structure correction occurs at $\alpha^4$. The Dirac treatment reproduces the terms at order 4 and yields additional terms at order 6 and above. 
According to the standard theoretical description, further corrections set in at $\alpha^5$ \cite{MohrNT2015}.

As presented in eqs.~\eqref{eq:FullGrdState} and \eqref{eq:FullTransFrequ}, we obtain the corrections induced by the T-duality propagator in terms of Bessel and Struve functions. When expanding these results in powers of the fine-structure constant, eqs.~\eqref{eq:seriesAlphaGrdState} and \eqref{eq:seriesAlphaTransFrequ}, we find the first manifestations at order $\alpha^4$ and $\alpha^4 \ln\left(1/\alpha\right)$. 
At second order in perturbation theory, we expect terms starting with $\alpha^5$. Similarly, mutual interactions with the fine-structure corrections can only appear at order 5 and above.
Therefore, we ensure to describe a proper observable since we take all contributions to order 4 into consideration. 

The other parameter which determines the amplitude of the T-duality induced correction is the zero-point length. For small $\lz$, we find an approximately quadratic dependency as evident in the series expansions eqs.~\eqref{eq:seriesGrdState} and \eqref{eq:seriesTransFrequ}. Furthermore, the logarithmic term is dominating and responsible for the overall sign of the corrections. Both these features are evident in Figs.~\ref{fig:plotConstrGrdStateWLogLogLegend} and \ref{fig:plotConstrFrequWLogLogLegend}. 

We stress that the fine-structure corrected Schr\"odinger value for the ground state energy lies below the experimental one. Also the current theoretical value hints at stronger binding than the current experimental one (cf.~Tab.~\ref{tab:ThBackground_Energy_Comparison}). Similarly, the analogous theoretical transition frequencies are found above the experimental counterpart (cf.~Tab.~\ref{tab:ThBackground_Frequ_Comparison}).
Thus, at qualitative level, a shift towards weaker binding energy and smaller transition frequency improves the match between the theoretical and experimental results.
Indeed, both hold true for the T-duality corrections.

At quantitative level, there are two ways of comparing potential theoretical contributions with experimental data. One can compare the conventional theoretical value with the experimentally measured one. Under the assumption that the discrepancy is generated by the novel effect only, one can derive a bound on the underlying parameter. Alternatively, one can focus on uncertainties and require the extra contribution to be smaller than the experimental accuracy. Then the experimental standard deviation is regarded as the reference scale. The latter approach usually results in stronger constraints and is used commonly in literature, e.g.~\cite{Brau1999,HossenfelderBHRSS2003}.
We follow both approaches. However, we refine the first way in a conservative manner: We do not just take into account the discrepancy between the theoretical and experimental value, but we also take into consideration the respective standard deviations. In this way, we obtain a less strict, but more reliable bound.

In this paper we address the ground state energy as well as the energy difference between the $1\inm{S}_{1/2}$ and $2\inm{S}_{1/2}$ level.
The latter observable turns out most suitable: Firstly, the corresponding transition frequency is known to a better precision, experimentally and theoretically. For instance, the experimental relative error exceeds the one of the ground state energy by 5 orders of magnitude.
Indeed, this transition frequency is the most accurately known hydrogen spectral line \cite{Kramida2010}.
Secondly, the difference between two levels is relative by definition: It is insensitive to global energy shifts. 

Overall, we find constraints in the range $\SI{1.6e-14}{\m}$ down to $\SI{3.9e-19}{\m}$. Table~\ref{tab:Constraints_Comparison} shows a compilation of the bounds obtained.
Limits in the range of $\SI{e-17}{\m}$ are also found in related minimal length considerations \cite{Brau1999,AntonacciOakesFFN2013} even though they arise from a different context: They rely on a generalization of the Heisenberg uncertainty relation while the limits presented here are the direct consequence of the T-duality propagator. 
There is a further sense of complementarity: They focus on a modification of the electron evolution and alter the Schr\"odinger or Dirac equation. In contrast, we applied the modified massless propagator to the electromagnetic field and studied the modified Coulomb potential while keeping the ordinary Schr\"odinger description. Although in both ways the energy contributions set in at $\alpha^4$, we find an additional logarithmic contribution.

More precise experimental and theoretical results would be helpful to further restrict the size of minimal lengths and to explore the viable range of quantum gravity modifications. Based on investigations of the smallest sensible size of black holes, the authors of \cite{NicoliniSW2019} expected the value of the zero-point length at $\lz = {(2/3)}^{3/4}\, \lpl \approx 0.8\, \lpl$ where $\lpl$ is the Planck length. Testing this length scale with the hydrogen atom would correspond to a relative precision of $\num{e-47}$ for the ground state as well as for the transition frequency. While this seems out of reach with the techniques present today, astrophysical observations of small black holes could provide further insights.

\section{Summary}
\label{sec:Summary}
String theory is a high-energy completion of quantum field theory and gravitation. Regarding quantum field excitations as strings instead of point-like particles, the authors of \cite{FontaniniSP2006} derived the modified 4-dimensional propagator. It introduces a parameter called zero-point length, $\lz$, which is closely related to the self-dual radius of T-duality. This approach was applied to the context of black holes and -- in principle observable -- differences to the general relativistic counterparts were identified \cite{NicoliniSW2019}. 

In this paper, we investigated how the modified propagator manifests itself in the hydrogen atom. The hydrogen atom is a well suited system since it is investigated to high precision in theory and experiment. We derived the corrections to the hydrogen ground state energy and the $1\inm{S}_{1/2} - 2\inm{S}_{1/2}$ transition frequency by first order Rayleigh-Schr\"odinger perturbation theory. Comparing with experimental data, we could derive constraints on the zero-point length ranging down to $\lz < \SI{3.9e-19}{\m}$.

\section*{Acknowledgements}
MFW gratefully acknowledges the support by the Stiftung Polytechnische Gesellschaft Frankfurt am Main. 
This research was supported in part by the Helmholtz International Center for FAIR within the framework of the LOEWE program (Landesoffensive zur Entwicklung Wissenschaftlich-\"{O}konomischer Exzellenz) launched by the State of Hesse.

\appendix
\section{Useful Identities}
\label{sec:Useful_Identities}
The energy shifts from the T-duality contribution presented in Sec.~\ref{subsec:T-Duality_Contribution} stem from non-standard integrals which result in special functions. The generalized form of these integrals results in
\begin{equation}
\int_0^\infty x^{2\nu-1}\, {\left(u^2 +x^2\right)}^{\rho-1}\, \eu^{-\mu x}\, \dext x
= \frac{u^{2\nu+2\rho-2}}{2\sqrt{\pi}\GamFct(1-\rho)}\; %
  \MeijerG*{3}{1}{1}{3}{1 -\nu}{1 -\rho -\nu,\, 0,\, 1/2}{\mu^2 u^2/4}
\end{equation}
for $\left|\arg u\right| < \pi/2$, $\Re \mu > 0$, and $\Re \nu > 0$ 
\cite[p.~351, eq.~3.389.2]{GradshteynR2007}. 
Here, 
$\MeijerG*{m}{n}{p}{q}{a_1,\ldots,a_p}{b_1,\ldots,b_q}{x}$
is the Meijer's $G$-function. 

The resulting expressions can be simplified by using representations of the special functions in terms of the Meijer's $G$-function. In particular, the Bessel functions of the second kind, $\YBessel{\nu}(x)$, and the Struve functions, $\HStruve{\nu}(x)$, turn out useful (\cite[p.~1034, eq.~9.34.2]{GradshteynR2007} and \cite[p.~1035, eq.~9.34.5]{GradshteynR2007}):
\begin{align}
\YBessel{\nu}(x)\, x^\alpha
&= 2^\alpha\, \MeijerG*{2}{0}{1}{3}{(\alpha-\nu-1)/2}{(\alpha-\nu)/2,\, (\alpha+\nu)/2,\, (\alpha-\nu-1)/2}{x^2/4}\\
\HStruve{\nu}(x)\, x^\alpha
&= 2^\alpha\, \MeijerG*{1}{1}{1}{3}{(\alpha+\nu+1)/2}{(\alpha+\nu+1)/2,\, (\alpha-\nu)/2,\, (\alpha+\nu)/2}{x^2/4}
\end{align}


\begingroup\raggedright

\endgroup

\end{document}